\documentclass[prd,onecolumn,unsortedaddress,superscriptaddress]{revtex4-2}
\usepackage[english]{babel}
\usepackage{graphicx}
\usepackage{epsfig,color}
\usepackage{amsmath,amssymb,eufrak,calc}
\usepackage{amsthm}
\usepackage{enumerate}
\usepackage{hyperref}
\usepackage{amsxtra,amsfonts,bm}
\usepackage{blkarray}
\usepackage{adjustbox}

\begin{document}

\title{Quantum Simulation of Lattice Gauge Theories in more than One Space Dimension - Requirements, Challenges, Methods }
\author{Erez Zohar}
\address{Racah Institute of Physics, The Hebrew University of Jerusalem, Givat Ram, Jerusalem 91904,  Israel.}

\begin{abstract}
Over the recent years, the relatively young field of quantum simulation of lattice gauge theories - aiming at implementing simulators of gauge theories with quantum platforms, has gone through a rapid development process. It is now of interest not only to people in the quantum information and technology community, but also seen as a valid tool for tackling hard, nonperturbative gauge theory physics by more and more particle and nuclear physicists. Along the theoretical progress, nowadays more and more experiments which actually implement such simulators are being reported, manifesting beautiful results, but mostly on $1+1$ dimensional physics. In this paper, we review the essential ingredients and requirements of lattice gauge theories in more dimensions, discuss their meanings, the challenges they impose and how they could be dealt with, potentially, aiming at the next steps of this field towards simulating challenging physical problems.
\end{abstract}

\maketitle

\section{Introduction}
Gauge theories are fundamental in many contexts of modern physics. \emph{Gauge invariance} is the manifestation of a local symmetry whose transformations act only on  a  small local set of degrees of freedom. Lifting a global symmetry of a matter theory to be local, usually  through the well-known \emph{minimal coupling} procedure, introduces \emph{gauge fields}. They account for the local symmetry and mediate the interactions between matter particles. Such theories may be found in the standard model of particle physics, including, for example, Quantum Electrodynamics (QED), with an Abelian gauge (symmetry) group,  or Quantum Chromodynamics, describing the strong nuclear force, with an $SU(3)$ symmetry.

While QED is weakly interacting and may be studied accurately using perturbation theory, QCD is very different. Due to its running coupling, in high energies limits it is very weakly coupled (asymptotically free), enabling perturbative treatment. Low energy QCD, however is a strongly interacting, non-perturbative theory. Among the important problems  of non-perturbative QCD is the \emph{confinement of quarks}, the responsible for binding quarks together into hadrons, whose breaking requires an infinite amount of energy, forbidding the existence of free quarks in the spectrum.

Back in the 1970s, lattice gauge theory (LGT) was proposed as an approach to deal with this problem \cite{wilson_confinement_1974}. What started as an analytical tool \cite{wilson_confinement_1974,kogut_hamiltonian_1975,polyakov_quark_1977} quickly emerged into an impressive numerical framework, in a combination with Monte-Carlo methods. That has given rise to many important QCD results and is still being used these days for static properties such as the hadronic spectrum \cite{aoki_flag_2020}. However, these computations are carried out in Wick-rotated Euclidean spaces, which blocks the road to direct obeservation of real-time dynamics as well as suffers from the sign problem in important finite chemical potential scenarios \cite{troyer_computational_2005}.

In the last decade, in parallel to the successful application of quantum information and technology based methods, such as quantum simulation \cite{feynman_simulating_1982} or tensor networks \cite{orus_practical_2014,cirac_matrix_2020} to many body models in condensed matter physics, the effort to generalise these methods to particle physics has begun \cite{wiese_ultracold_2013,zohar_quantum_2016,dalmonte_lattice_2016,preskill_simulating_2018,banuls_simulating_2020,banuls_review_2020}. Quantum simulation of LGTs, their mapping into controllable quantum platforms that are controllable in the lab, allowing one to experimentally study inaccessible physics, has evolved from the early theoretical proposals of analog simulators \cite{zohar_confinement_2011,banerjee_atomic_2012}, digital ones \cite{buchler_atomic_2005,tagliacozzo_optical_2013} or quantum computer algorithms \cite{byrnes_simulating_2006} into an experimental field \cite{martinez_real-time_2016,kokail_self_2019,schweizer_floquet_2019,mil_scalable_2020,yang_observation_2020}, attracting particle and nuclear physicists \cite{cloet_opportunities_2019} who now consider it to be a valid computational method.

In this article we will address the challenges faced by the quantum simulation of LGTs, from the theoretical point of view, in particular in more than one space dimension (which has so far been less explored experimentally). In the following sections, we will provide some basic introduction to the theories in question, formulate a list of requirements which must be satisfied, bypassed or overcome in such simulators and discuss them.

 \section{Degrees of freedom, Hilbert Space, Symmetry}
 
 When discussing the quantum simulation of a physical model, a proper map needs to be devised between the simulated model and the simulating platform. This map can be exact (in the ideal, almost eutopic case) or approximate (with proper error bounds and accuracy estimations). The simulation can be \emph{analogue}, involving a mapping of the Hamiltonian; \emph{stroboscopic} or analogue-digital, in which different parts of the Hamiltonian are implemented sequentially, using short time pulses similar to quantum gates, based on the Trotter-Suzuki expansion \cite{suzuki_decomposition_1985} and possibly involving ancillary degrees of freedom; or completely \emph{digital}, implementing the dynamics via a non-universal quantum circuit. But what all these approaches share is that one needs the simulator to encode the degrees of freedom, the Hilbert spaces, the symmetries and the interactions of the simulated model, whatever choice is made for the simulation approach. Those ingredients of the simulated model thus need to be carefully examined, before one can practically reconstruct the model using other building blocks. And the closer the available elementary blocks and interactions of the simulating platform are to those of the simulated model, the better the quantum simulator is. These arguments are fundamental when one considers the design of a quantum simulator, and they lead to proper choices of the simulating platform, the simulation approach and the mapping method. We will thus begin our discussion with a general review of the ingredients of a lattice gauge theory and the requirements they impose on their quantum simulation - in particular in more than a single space dimension. 
  
Following the Hamiltonian approach to LGTs \cite{kogut_hamiltonian_1975}, we only discretize space, and time is a continuous, real coordinate. The lattice sites, hosting the matter fields (usually fermionic, but can be bosonic too) are labelled by $\mathbf{x}\in\mathbb{Z}^d$. $\left\{\mathbf{e}_i\right\}_{i=1}^d$, are unit vectors in the positive directions. The links, which hosts the gauge field degrees of freedom, are denoted by pairs of a starting site and a direction, $\left(\mathbf{x},i\right)$. 

This immediately brings us to the first requirement from an LGT quantum simulator: it should include two different types of degrees of freedom - gauge field and matter, residing on the links and the sites, respectively. In the most general settings, one needs to consider a quantum simulation platform which enables the interplay of fermionic and non-fermionic degrees of freedom - e.g. ultracold atoms in optical lattices \cite{jaksch_cold_1998}. Other platforms without fermions (e.g. trapped ions or superconducting qubits) may be used (i) 
when pure gauge theories (without matter) are simulated,  or when the matter is bosonic; (ii) in a single space dimension, where fermionic 
matter may be mapped conveniently to spins, using the Jordan-Wigner map \cite{jordan_uber_1928}, such as in \cite{martinez_real-time_2016,yang_observation_2020,davoudi_towards_2020,atas_su_2021}); (iii) when the fermionic matter is mapped, using the gauge field which allows that in some cases, to hard-core bosonic matter \cite{zohar_eliminating_2018} (this can be done locally and unitarily, and in some cases the matter may be completely eliminated \cite{zohar_removing_2019}). Finally, one may also use only fermionic degrees of freedom, when constructing all the gauge field operators by auxiliary fermionic modes residing on the links \cite{brower_qcd_1999}, as used e.g. in \cite{banerjee_atomic_2012,banerjee_atomic_2013}.

Gauge transformations are in the centre of gauge theories. They are parametrised by elements of the \emph{gauge group} $G$, which is usually a compact Lie group (in particle physics scenarios) or a finite one (in several condensed matter contexts). On each site $\mathbf{x}$ one defines the unitary gauge transformation operators $\hat{\Theta}_g\left(\mathbf{x}\right)$ for each $g \in G$, acting only on a finite volume of space - the matter degree of freedom at $\mathbf{x}$, and the gauge fields on the links intersecting in it.  A state $\left|\psi\right\rangle$ is gauge invariant if
\begin{equation}
	\hat{\Theta}_g\left(\mathbf{x}\right) \left|\psi\right\rangle = \left|\psi\right\rangle \quad \quad \forall \mathbf{x} \in \mathbb{Z}^d , g \in G
	\label{psitrans}
\end{equation}
(but not only if; we disregard here the case of \emph{static charges}, a discussion of which may be found in \cite{kasper_from_2020}). An an operator $O$ is gauge invariant if and only if
\begin{equation}
	\hat{\Theta}_g\left(\mathbf{x}\right) O 	\hat{\Theta}^{\dagger}_g\left(\mathbf{x}\right) = O  \quad \quad \forall \mathbf{x} \in \mathbb{Z}^d , g \in G
	\label{Otrans}
\end{equation}

This allows us to formulate the second requirement: Quantum Simulators of lattice gauge theories should be gauge invariant - that is, manifest a local symmetry parametrised by the gauge group $G$. Gauge transformations $\hat{\Theta}_g\left(\mathbf{x}\right)$ should be well defined on each site $\mathbf{x} \in \mathbb{Z}^d$ and for any $g \in G$; the system has to be prepared initially in a gauge invariant state $\left|\psi\right\rangle$ as in Eq. (\ref{psitrans}); and the dynamics should include gauge invariant interactions. Normally, such symmetries are not available for free in the simulating platforms, and one has to impose them or map them to an existing symmetry of the simulating platform, if possible. 

To discuss all that more concretely, let us zoom into the properties of the matter and gauge field Hilbert spaces.

\section{Matter Fields}
The matter degrees of freedom in most LGTs, especially in particle physics context, are fermionic. In which case, on each site $\mathbf{x}$ we introduce a set of fermionic creation operators, $\psi^{\dagger}_m\left(\mathbf{x}\right)$. In general, the indices carried by the fermionic operators may label spin (related to the Lorentz group), flavour or colour (related to the gauge symmetry). There are different methods for dealing with fermions on a lattice, having to do with the doubling problem \cite{kogut_lattice_1983} which we shall not discuss here. In some LGTs, however, bosonic matter or first-quantised, spin-like matter, may be considered as well.
For simplicity, in many quantum simulation works the staggered fermions formulation \cite{susskind_lattice_1977} is used: the spin components are distributed over several lattice sites,  recombined into spinors in the continuum limit. In this case, less fermionic species need to be considered on each site, which eases the technological demands from the simulator. Nevertheless, works on simulation of other fermionic formulations, such as naive \cite{zohar_simulating_2013} or Wilson fermions \cite{zache_quantun_2018}.

Flavour indices may be introduced when multi-flavour physics is studied, and in most cases this may only have experimental consequences having to do with the inclusion of more fermions and making sure that they all interact in the desired way; from the theoretical point of view of constructing a model or its simulation, the generalisation from one flavour to more is straight-forward and hence will not be discussed here. Finally, we are left with the colour indices, having to do with the gauge symmetry.

The modes $\psi^{\dagger}_m\left(\mathbf{x}\right)$ on each site form a complete spinor in a given representation of the gauge group $G$, mostly an irreducible one (irrep). That implies that when Abelian groups, such as $U(1)$ and $\mathbb{Z}_N$ are considered, we will only have one fermionic mode per site, $\psi^{\dagger}\left(\mathbf{x}\right)$. In non-Abelian cases, this spinor will correspond to a higher-dimensional irreducible representation - e.g. the fundamental one with dimension $N$ (and hence $N$ fermionic components) for $SU(N)$ or $U(N)$. Let us restrict ourselves to this choice, for simplicity; we then define local unitary transformations  $\theta_g\left(\mathbf{x}\right)$ parametrised by the group elements, for which
\begin{equation}
\theta_g\left(\mathbf{x}\right) \psi^{\dagger}_m\left(\mathbf{x}\right) \theta_g^{\dagger}\left(\mathbf{x}\right) =
\psi^{\dagger}_{m'}\left(\mathbf{x}\right)D_{m'm}\left(g\right)
\end{equation}
where $D\left(g\right)$ is the unitary representation of $g$ in the irreducible representation picked for the matter. When $G$ is a compact Lie group, its elements $g$ are completely determined in terms of group parameters, or coordinates on the group manifold, $\phi_a\left(g\right)$. Then, we may use the generators (local charge operators) $Q_a\left(\mathbf{x}\right)$, satisfying the group's algebra
$
	\left[Q_a\left(\mathbf{x}\right),Q_b\left(\mathbf{y}\right)\right]=i\delta\left(\mathbf{x},\mathbf{y}\right)f_{abc}Q_a\left(\mathbf{x}\right)
$,
where $f_{abc}$ are the group's structure constants, to express the transformation,
$
	\theta_g\left(\mathbf{x}\right) = \exp\left(i \phi_a\left(g\right) Q_a\left(\mathbf{x}\right)\right)
$.
In the case of fermionic matter, we can express the local charges as bilinears of the mode operators, e.g.
$
	Q_a\left(\mathbf{x}\right) = \psi^{\dagger}_m \left(\mathbf{x}\right) \left(T_a\right)_{mn}	
	\psi_n \left(\mathbf{x}\right)
$
, where $T_a$ are the suitable matrix representations of the generators, with
$
	\left[T_a,T_b\right]=if_{abc}T_c
$
(with some modifications depending on the fermionic prescription used, for example see \cite{zohar_formulation_2015} for the staggered case).

\section{Gauge Fields}
In the continuum, gauge fields are bosonic; their excitations are countable particles, such as photons and gluons, which are described within bosonic Fock spaces. On the lattice, however, the gauge fields are rather \emph{first quantised} objects, described in terms of simple Hilbert spaces on the links, rather than Fock spaces with particle excitations. Let us see that. 
 On each link of the lattice, the gauge field Hilbert space may be spanned in terms of two convenient bases (among others). We shall postpone the discussion of one of those, \emph{the representation basis}, to later, and introduce now the other one - the \emph{group element basis}. As implied by its name, the elements of this basis are labelled by elements of the gauge group - $\left\{\left|g\right\rangle\right\}_{g\in G}$. When $G$ is finite, we denote by $\left|G\right|$ the number of its elements, and this is the \emph{finite} dimension of the Hilbert space we have on each link (for example, when $G=\mathbb{Z}_N$, the group elements may be identified with the $N$th roots of $1$, and $\left|G\right|=N$). The basis states then satisfy the orthonormality relation
\begin{equation}
	\left\langle g | h\right\rangle = \delta_{gh}
	\label{orth}
\end{equation}
with the Kronecker delta. 
When $G$ is a compact Lie group, $\left|G\right|$ denotes the group's volume, defined by $\left|G\right|=\int dg$ - integration of the group's Haar measure $dg$ over the whole manifold; $\delta_{gh}$ will hence be a distribution.

We realise that the on-link gauge field Hilbert spaces of a lattice gauge theory are first quantised, and do not require the use of any second-quantised, Fock space objects. However, in the case of infinite groups (e.g. compact Lie ones), these Hilbert spaces are infinite, which might be problematic to implement on the simulation platforms. We will discuss truncation schemes below.

Since each link is connected to two sites, on which two differend, independent gauge transformations may be defined, the gauge field on a link can be affected by two independendent gauge transformations, associated with either of its ends. We thus define unitary operators $\Theta_g\left(\mathbf{x},i\right)$ and $\tilde\Theta_g\left(\mathbf{x},i\right)$ acting on the gauge field on the $\left(\mathbf{x},i\right)$ link, transforming it from the right and left sides respectively; in the group element basis, they have a simple interpretation as group multiplication operators,
\begin{equation}
	\Theta_g \left|h\right\rangle = \left|hg^{-1}\right\rangle  \quad \text{ and } \quad \tilde\Theta_g \left|h\right\rangle = \left|g^{-1}h\right\rangle  \quad \forall g,h \in G
\end{equation}
We also define unitary transformation opeartors for the matter degrees of freedom on the sites, $\theta_g\left(\mathbf{x}\right)$. Its explicit definition depends on the nature of the matter. Altogether, the local gauge transformation at $\mathbf{x}$ parametrised by $g$ is implemented with the unitary
\begin{equation}
	\hat{\Theta}_g\left(\mathbf{x}\right) = \overset{d}{\underset{i=1}{\prod}}\left[\tilde\Theta_g\left(\mathbf{x},i\right)\Theta^{\dagger}_g\left(\mathbf{x},i\right)\right]\theta^{\dagger}_g\left(\mathbf{x}\right)
	\label{Gtrans}
\end{equation}

When $G$ is a compact Lie group, we can express the operators $\Theta_g$ and $\tilde\Theta_g$ using generators and group parameters. Let us introduce two sets of generators, $\left\{R_a\right\}$ and $\left\{L_a\right\}$, generating the right and left transformations respectively:
\begin{equation}
	\Theta_g = e^{i\phi_aR_a}; \quad \tilde\Theta_g = e^{i\phi_aL_a}
\end{equation}
where $\left\{\phi_a\left(g\right)\right\}$ is the set of group parameters (coordinates) uniquely associated with $g$, and the generators satisfy the group's mutually indepdnedent right and left Lie algebras,
$
	\left[R_a,R_b\right]=if_{abc}R_c;$ $\left[L_a,L_b\right]=-if_{abc}L_c ;$ and $\left[R_a,L_b\right]=0$.
When $G$ is Abelian, left and right transformations are the same, all the generators commute and left and right transformations are the same. In the $U(1)$ case, for example, there is only one generator, $E=R=L$. In the $SU(2)$ case, there are two sets of "angular momentum operators", right and left, which in the rigid body picture may be seen as the angular momenta in the space and body frames \cite{kogut_hamiltonian_1975,kasper_from_2020}. 

Hence, in the compact Lie case, the local gauge transformations are generated by the operators
\begin{equation}
	G_a\left(\mathbf{x}\right) = \overset{d}{\underset{i=1}{\sum}}\left(L_a\left(\mathbf{x},i\right)-R_a\left(\mathbf{x}-\mathbf{e}_i,i\right)\right)-Q_a\left(\mathbf{x}\right)
\end{equation}
and gauge invariant states satisfy
\begin{equation}
	G_a\left(\mathbf{x}\right)\left|\psi\right\rangle = 0 \quad \forall \mathbf{x} \in \mathbb{Z}^d, a
	\label{Gausslaw}
\end{equation}
(disregarding static charges again).
In the $U(1)$ case, we have $G\left(\mathbf{x}\right) = \overset{d}{\underset{i=1}{\sum}}\left(E\left(\mathbf{x},i\right)-E\left(\mathbf{x}-\mathbf{e}_i,i\right)\right) - Q\left(\mathbf{x}\right)$, and thus 
\begin{equation}
	\overset{d}{\underset{i=1}{\sum}}\left(E\left(\mathbf{x},i\right)-E\left(\mathbf{x}-\mathbf{e}_i,i\right)\right)\left|\psi\right\rangle = Q\left(\mathbf{x}\right)\left|\psi\right\rangle
	\label{U1gauss}
\end{equation}
- the divergence of the gauge field operator $E$ on a site equals the charge  there $Q\left(\mathbf{x}\right)$. With this we immediately recognise $E$ as the \emph{electric field} operator on the link, and Eq. (\ref{U1gauss}) as the lattice \emph{Gauss law}; generalising to non-Abelian groups, we call $R_a$ and $L_a$ the right and left electric fields, respectively, and then (\ref{Gausslaw}) is the rather more general, non-Abliean form of the Gauss law.

\subsection{Imposing Gauge Invariance on a Quantum Simulator}
	 So far, several methods have been proposed for the imposition of gauge invariance in quantum simulators. The first is \emph{effective gauge invariance} - an analogue scheme, in which the desired gauge-invariant Hamiltonian is only implemented effectively. One implements, instead, a different, so-called \emph{primitive Hamiltonian} $H'$, violating the local symmetry ($\left[H',\hat{\Theta}_g\left(\mathbf{x}\right)\right] \neq 0$), but including penalty terms which constrain the symmetry. 
For example, let us consider the first analogue schemes for Kogut-Susskind simulators \cite{zohar_confinement_2011,banerjee_atomic_2012,zohar_simulating_2012}, which discussed cold atomic implementations of a $U(1)$ theory (compact QED). It is an Abelian model whose gauge transformations are generated by a single operator $G\left(\mathbf{x}\right)$ on each site: $\hat{\Theta}_{\phi}\left(\mathbf{x}\right) = e^{i \phi G\left(\mathbf{x}\right)}$ for every $\phi \in \left[0,2\pi\right)$. As in (\ref{Gausslaw}), for a gauge invariant state $\left|\psi\right\rangle$,  $G\left(\mathbf{x}\right)\left|\psi\right\rangle=0$ on each site $\mathbf{x}$. If $H'$ contains  penalty terms of the form $\lambda\underset{\mathbf{x}}{\sum}G^2\left(\mathbf{x}\right)$, and $\lambda$ is significantly larger than any other energy scale in $H'$,  the spectrum will be split into sectors corresponding to eigenvalues of $G\left(\mathbf{x}\right)$. One can perturbatively derive an effective Hamiltonian not mixing these sectors, assuming that transitions between them are only possible virtually. As a result, in the lowest energy sector we will get an effective Hamiltonian with the local symmetry we wanted.

Usually, when the interactions in the primitive Hamiltonian are properly tailored, the second order is enough for the desired interactions (and the remaining, higher terms are still gauge invariant). The method, which can be extended to non-Abelian cases  \cite{banerjee_atomic_2013}, is considered slightly forced and less robust than others, since it is sensitive to experimental imperfections and requires very careful fine-tuning, especially in the non-Abelian cases where there are several $G_a$ operators on each site (e.g. three for $SU(2)$). Furthermore, although the $G_a\left(\mathbf{x}\right)$ operators are local, squaring them introduces interactions which may be hard to implement. More recently, however, it was shown that one may also use local constraints which are linear in the generators  - that is,  without interactions, in both the Abelian \cite{halimeh_gauge_2020} and non-Abelian (using dynamical decoupling) \cite{kasper_non_2020} cases. Another option to introduce constraints linearly is by using dissipation and non-unitary dynamics, utilising the Zeno effect to force gauge invariance \cite{stannigel_constrained_2014}.

Another analogue approach is \emph{symmetry mapping.} There, the local symmetry is guaranteed in  a more fundamental level. One maps (when possible) a symmetry of the simulating platform to the local symmetry of the simulated model. It was introduced in a proposal for simulating lattice gauge theories with ultracold atoms in optical lattices \cite{zohar_quantum_2013}, where the atomic states are classified by their hyperfine angular momentum, and the available interactions are two-atom collisions. The collision preserves the total hyperfine angular momentum, and thus, by properly choosing the hyperfine states playing different roles in the simulator - fermions on the sites for the matter and bosons, on the links (out of which the gauge field operators are constructed through through Schwinger representations), one may exactly map all the gauge invariant processes to angular momentum conserving collisions. This was further analysed  \cite{kasper_implementing_2017} and demonstrated experimentally by implementing a single building block \cite{mil_scalable_2020}. In another recent example for symmetry mapping, a local symmetry of a Rydberg atom state was exactly mapped to gauge invariance \cite{surace_lattice_2020}.

Finally, \emph{digital and stroboscopic implementations} offer various ways to impose gauge invariance: when the simulation does not involve a mapping of the Hamiltonian but rather an implementation of the time evolution by short time pulses implementing different interactions and Hamiltonian terms separately (trotterisation), possibly using auxiliary degrees of freedom, there are more options to enforce the symmetry. These range from tailoring gauge invariant interactions with an ancilla \cite{tagliacozzo_optical_2013,tagliacozzo_simulation_2012,zohar_digital_2017,zohar_digital_2017-1,bender_digital_2018,lamm_general_2019} to developing tools to encode the symmetries on quantum computer algorithms \cite{byrnes_simulating_2006,klco_quantum_2018,klco_su_2020,raychowdhury_loop_2020,ciavarella_trailhead_2021}. When such methods are used, it is important to monitor gauge invariance and its possible breaking. In some each step of the strobscopic dynamics preserves the symmetry \cite{zohar_digital_2017-1} and thus  trotterisation  does not affect the gauge invariance. More generally, some works have been written on monitoring gauge invariance in digital simulations, as in   \cite{stryker_oracles_2019}. 

One may also deal with gauge invariance in the simulation, by completely bypassing it. The gauge symmetry implies some redundancy in the description of the physical state. To lift it one can use the local constraints to eliminate the matter or the gauge field degrees of freedom. Eliminating the gauge field degrees of freedom by solving the Gauss law is possible in a single space dimension \cite{martinez_real-time_2016}, but not in further dimensions. Eliminating the matter fields is possible, in some cases, in any dimension \cite{zohar_removing_2019}. The other way to lift the redudancy is to change to a dual formulation of gauge theories, with \emph{magnetic} variables, where there is no local symmetry anymore. Such formulations were introduced in the pure gauge case \cite{drell_quantum_1979,kaplan_gauss_2020,celi_emerging_2020} as well as in the presence of dynamical fermionic matter, using translational invariant, Green's function based approach \cite{bender_gauge_2020} or a maximal tree approach \cite{haase_resource_2020}.

\subsection{Gauge Field Bases}

As discussed, the gauge field Hilbert space on a link may be spanned by the group element basis states, $\left\{\left|g\right\rangle\right\}_{g\in G}$. These are the eigenstates of the group element operators, $U^j_{mn}$. Let $j$ be an irreducible representation of $G$, and $D^j\left(g\right)$ the $j$ unitary representation of the group element $g$. Then, we define
\begin{equation}
	U^j_{mn} = \int dg D^j_{mn}\left(g\right) \left|g \right\rangle\left\langle g \right|
\end{equation}
(where the integral is replaced by a sum if the group is finite). This is a matrix of operators acting on the link's Hilbert space, all commuting with one another, as can clearly be seen in this definition: they are all diagonal in the group element basis. It is straightforward to see that $\Theta_g U^j_{mn} \Theta^{\dagger}_g = U^j_{mn'}D^j_{n'n}\left(g\right)$ as well as  $\tilde \Theta_g U^j_{mn} \tilde \Theta^{\dagger}_g = D^j_{mm'}\left(g\right)U^j_{m'n}$ - the group element operator transforms like a unitary matrix representation of $G$; in the Lie group case these are equivalent to
$\left[R_a,U^j_{mn}\right] = U^j_{mn'}\left(T^j_a\right)_{n'n}$ and $\left[L_a,U^j_{mn}\right] = \left(T^j_a\right)_{mm'}U^j_{m'n}$, respectively. In the Abelian, $U(1)$ case, all the irreps are one dimensional, with $U^j = e^{ij\phi}$ for any integer $j$; the operator $\phi$ plays the role of the compact vector potential, and $\left[E,U\right]=U$ implies that it is canonically conjugate to $E$.

Another useful basis is the \emph{representation basis}, whose elements are eigenstates of the group transformation operators $\Theta_g$ and $\tilde\Theta_g$, or more specifically - of a maximal set of mutually commuting operators thereof. Its elements, $\left|jmn\right\rangle$, are obtainable from the group element states through 
\begin{equation}
\left\langle g|jmn\right\rangle = \sqrt{\frac{\text{dim}\left(j\right)}{\left|G\right|}}D^j_{mn}\left(g\right)
\end{equation}
where $j$ labels an irreducible representation, with dimension $\text{dim}\left(j\right)$. In the case of a compact Lie group, one can recognise $j$ as the eigenvalue(s) of Casimir operator(s), and $m,n$ are, respectively, eigenvalues of the maximal set of mutually commuting $L_a$ and $R_a$ operators. When the group is abelian, all the irreducible representations are one dimensional, and hence there are no $m,n$ quantum numbers, and the representation states are eigenstates of the electric field operator, with integer spectrum $E\left|j\right\rangle = j\left|j\right\rangle$ (the Hilbert space of a particle on a ring, with coordinate $\phi$ and angular momentum $E$). The $SU(2)$ link Hilbert space is that of a rigid rotator \cite{kogut_hamiltonian_1975,kasper_from_2020}: the coordinate (group element) state may be parametrised by the three Euler angles, while the representation basis is given by angular momentum operators, responsible for rotations in the space ($R_a$) and body ($L_a$) frames: commuting rotations giving rise to the same total angular momentum (rotation scalar) $\mathbf{J}^2 = R_aR_a = L_aL_a$. Then we have $\mathbf{J}^2\left|jmn\right\rangle = j\left(j+1\right) \left|jmn\right\rangle$, 
$L_z\left|jmn\right\rangle = m \left|jmn\right\rangle$ and $R_z\left|jmn\right\rangle = n \left|jmn\right\rangle$.

Generally speaking, in the non-Abelian case one can formally decompose the representation state into a product of a left state and a right state, $\left|jmn\right\rangle = \left|jm\right\rangle \otimes \left|jn\right\rangle$ carrying the same representation. This implies that although it may be tempting to divide the link of a lattice gauge theory into two different parts, the left and the right ones,  trasformed by the gauge transformations of the link's beginning and end, respectively, the two parts are not independent, and are constrained by having the same irrep $j$. Formally, the Hilbert space on a link may be written as
\begin{equation}
	\mathcal{H}_{\text{link}} = \underset{j}{\bigoplus}\left(\mathcal{H}_j\otimes\mathcal{H}_j\right)
	\neq \left(\underset{j}{\bigoplus}\mathcal{H}_j\right) \otimes \left(\underset{j'}{\bigoplus}\mathcal{H}_{j'}\right)
\end{equation}
(where $\mathcal{H}_j$ is the Hilbert space spanned by the multiplet $\left|jm\right\rangle$, with a fixed $j$) and one cannot simply decompose the link Hilbert space into a product of two (this is also manifested through the group element basis, which takes the link as a whole). There are several formulations of LGTs which do break the link into two (which may have exactly the advantage of coupling to two different Gauss laws) but then a constraint is imposed. Thus, when used for quantum simulation, one must make sure that the constraint between the two parts of each link is not violated, for example through introducing a penalty term forcing it into the Hamiltonian. This is done, for example, in the link model \cite{brower_qcd_1999}, used for several quantum simulation schemes (\cite{banerjee_atomic_2012,banerjee_atomic_2013,wiese_ultracold_2013} and more) or the prepotential approach \cite{mathur_loop_2007}, used for example in \cite{zohar_cold-atom_2013} and \cite{raychowdhury_loop_2020}. In another approach \cite{zohar_formulation_2015} the link is kept as a whole.

\subsection{Gauge Field Truncations}
Another possible issue when designing quantum simulators of LGTs has to do with the dimension of the link Hilbert spaces: when the groups are infinite, as in the compact Lie case, these Hilbert spaces are infinite, which may be somewhere between inconvenient and impossible for common quantum simulation platforms. Then, one can truncate the gauge field Hilbert spaces in a way that simplifies this problem - but, as usual, may raise other ones.

When truncating the gauge field Hilbert spaces, we keep only a finite number of states, either in the group element or representation basis. Each truncation has its own advantages and disadvantages, depending on the simulation scheme and goals. In the representation basis, it is easy to put a cutoff on the representations we use. It follows from the Clebsch-Gordan series that, in the representation basis, the group element operator takes the form
$
U^j_{mn} = \underset{J,K}{\sum}\sqrt{\frac{\text{dim}\left(J\right)}{\text{dim}\left(K\right)}}
\left\langle J M j m | K M \right\rangle
\left\langle K N | J N j n \right\rangle
$
which suggests an immediate truncation scheme \cite{zohar_formulation_2015}, in which we restrict our Hilbert space only to representation states of some representations, connected by nonvanishing Clebsch-Gordan coefficients. For example, in $SU(2)$ the smallest truncation is five dimensional, with the representations $j=0$ ($\left|000\right\rangle$) and $j=\frac{1}{2}$ (with ($\left|\frac{1}{2},\pm\frac{1}{2},\pm\frac{1}{2}\right\rangle$). The great advantage of such a truncation is that although the Hilbert space is truncated, the gauge group remains the original one: the truncated $U$ operator still has the same transformation properties under $\Theta_g$ and $\tilde\Theta_g$, for every $g \in G$. This is very useful when one wishes to impose the symmetry using penalty terms, with the original Gauss laws, or use the symmetry mapping techniques. 

On the other hand, since we truncated the representation basis, its dual basis - that of group elements - cannot be used (and is ill defined) after the truncation, and one cannot use group elements to describe the Hilbert space any longer. Furthermore, the truncated group element operator is no longer unitary as a matrix, and its elements are no longer commuting operators, which blocks the way to some simulation schemes, such as the stroboscopic stator method \cite{zohar_digital_2017-1}, where these properties are crucial. 

As for the physical implications of this truncations - they are not yet clear in all the physical scenarios and theories. The full Hilbert space limit is obviously obtained when the cutoff is raised until all the representations are included again; one may argue that the truncations only affect short range physics but become less and less significant when large scale properties are sought and blocking is considered;  and in the link model formulation \cite{brower_qcd_1999} one may introduce an extra compact dimension which, once integrated over properly, allows to obtain the continuum the full gauge theory in the continuum limit from a truncated model on the lattice. In any case, however, this truncation scheme allows one to implement physical models with the full gauge group, which are of great interest and exhibit rich physics even without having a proper continuum limit: this opens the way to the study of many interesting lattice gauge theories that have not been considered before \cite{banerjee_atomic_2012,tagliacozzo_simulation_2012,wiese_ultracold_2013,banerjee_atomic_2013} and now are within experimental reach \cite{surace_lattice_2020}.

Another option is to truncate in the group element basis: choosing a subset of group elements instead of the full group. When this subset is a subgroup, nice properties such as $U_{mn}$  being a unitary matrix and commutation of all its elements as Hilbert space operators are kept; however the gauge group is narrowed to a finite subgroup and the Gauss laws no longer exist, preventing for example the option of Gauss law penalty terms. This is very useful, however, when using the stator scheme \cite{zohar_digital_2017-1}.

The most ideal scenario for such a truncation would be to identify an infinite series of finite subgroups of our gauge groups, which converge to the full group. Then we could think of a series of quantum simulators, all exhibiting gauge invariance, converging to the full model whose results could be deduced via extrapolation techniques. While this is the case for $U(1)$, which is the infinite $N$ limit of $\mathbb{Z}_N$, this is not the case for other relevant groups, such as $SU(2)$ or $SO(3)$: the crystallographic restriction tells us that the number of relevant finite transformation groups is finite (and very small). Then, of course, a due question is whether we are actually interested in implementing some gauge symmetry or could think of a simulation scheme in which some group elements, not necessarily forming a group, are chosen for the simulation.

\section{The Hamiltonian}

The last important ingredient to consider is the Hamiltonian. Normally,  a lattice gauge theory is described by the Kogut-Susskind Hamiltonian \cite{kogut_hamiltonian_1975}, consisting of four parts - two free and two interacting.

The free parts include (i) the matter mass term, depending on local number operators. For example, in the case of staggered fermions in $d$ space dimensions, it will have the form
\begin{equation}
	H_M = M\underset{\mathbf{x}}{\sum}\left(-1\right)^{x_1 + ... + x_d}
	 \psi^{\dagger}_m\left(\mathbf{x}\right)  \psi_m\left(\mathbf{x}\right)
\end{equation}
and (ii) the electric energy term, consisting of projectors onto the representations of the gauge field states on the link,
\begin{equation}
	H_E = \frac{g^2}{2}\underset{\mathbf{x};i=1,...,d}{\sum}\underset{j}{\sum}f_j\left|jmn\right\rangle\left\langle jmn \right|_{\mathbf{x},i}
\end{equation}
where $g$ is the coupling constant.
In the compact Lie group case it will depend on the quadratic Casimir operator, 
$	H_E = \frac{g^2}{2}\underset{\mathbf{x};i=1,...,d}{\sum}\mathbf{J}^2\left(\mathbf{x},i\right)$ (simply $E^2$ for $U(1)$) - resembling the well known continuum expression for electric energies. 

The interacting parts include (i) the link interaction - hopping of fermions to neighbouring sites, minimally coupled to the gauge field on the link in the middle, ensuring the gauge symmetry. For example, it will take the form
\begin{equation}
H_{\text{int}}=\underset{\mathbf{x},i}{\sum}\left(\epsilon_i\left(\mathbf{x}\right)\psi^{\dagger}_m\left(\mathbf{x}\right)U_{mn}\left(\mathbf{x},i\right)\psi^{\dagger}_n\left(\mathbf{x}+\hat{e}_i\right) +h.c.\right)
\end{equation}
(where the tunneling amplitudes $\epsilon_i\left(\mathbf{x}\right)$ are chosen in a way the produces the right fermionic dispersion relation \cite{susskind_lattice_1977}), and the representation chosen for the group element operator is the same as that of the matter. Finally, we have (iv) the magnetic energy term, which is a four body interaction of all the links around each plaquette of the lattice. It is a complicated interaction term, which becomes free in the continuum limit (corresponding to the magnetic field squared, and hence the name magnetic energy) and does not exist in a single space dimension. It takes the form
\begin{equation}
	H_B = -\frac{1}{2g^2}
	\underset{\mathbf{x};i<j}{\sum}\left(\text{Tr}
	\left(
	U\left(\mathbf{x},i\right)
	U\left(\mathbf{x}+\hat{e}_i,j\right)
	U^{\dagger}\left(\mathbf{x}+\hat{e}_j,i\right)
	U^{\dagger}\left(\mathbf{x},j\right) + h.c.
	\right)\right)
	\end{equation}
where the trace is over the matrix indices of the group element operators, and is not needed in the Abelian cases.

This plaquette interaction is one of the biggest challenge of quantum simulators of LGTs in more than one space dimension. Such plaquette or ring exchange interactions were considered for condensed matter gauge theories in the past, e.g. in \cite{buchler_atomic_2005}. In the context of Kogut-Susskind Hamiltonian, there are several ways to implement them, which have not been implemented experimentally so far, as they are either complicated, or involve weak interactions, or both. Possible analogue approaches include implementing the plaquettes when the Gauss law is added as a constraint, and the primitive Hamiltonian, as explained earlier in this work, is chosen propely \cite{zohar_confinement_2011,zohar_simulating_2012}. Other types of approaches include the motion of virtual particles, that cannot move physically, around a plaquette, carrying with them the $U$ fluxes around a plaquette and closing the trace. This is useful (but gives rise to weak interactions) in the symmetry mapping approach \cite{zohar_quantum_2013}. Another approach based on the effective motion of particles closing a plaquette was recently introduce \cite{homeier_Z2_2020}.

On the stroboscopic or digital side, one may used trotterised time evolution to build the plaquette interactions sequentially, as was shown first for pure gauge models, both Abelian and non-Abelian \cite{tagliacozzo_optical_2013,tagliacozzo_simulation_2012}. Such a stroboscopic scheme, in which the time evolution is trotterised and the plaquette interaction is generated through sequences of two-body interactions with ancillary degrees of freedom, and matter may be simulated as well, was introduced in \cite{zohar_digital_2017,zohar_digital_2017-1}, based on objects called stators \cite{zohar_half_2017}. One of the advantages of this scheme is that the stroboscopic evolution does not affect the gauge symmetry and it is completely protected \cite{zohar_digital_2017-1,bender_gauge_2020}.

Another way to deal with the plaquette interactions is to simply avoid them by switching to the dual picture, when available. There, the magnetic Hamiltonian becomes local, and the electric Hamiltonian becomes interacting - and only with two-body interactions \cite{drell_quantum_1979,kaplan_gauss_2020}. This can be very useful for quantum simulation of pure gauge theories \cite{celi_emerging_2020}, but when dynamical matter is introduced \cite{bender_gauge_2020,haase_resource_2020,paulson_towards_2020}, other parts of the Hamiltonian become complicated and non-local, but tradeoffs have always been and will most likely always be central to quantum simulation.

\section{Summary}

We have reviewed the challenges faced by the quantum simulation of LGTs in more than one space dimension, from a theoretical point of view, which can be summarizes in the following list of requirements (which have been discussed throughout the paper, including ways to bypass some of them):

\begin{enumerate}
	\item \textbf{Quantum simulators of LGTs with fermionic matter in more than $1+1d$ must allow for a way to describe the fermionic statistics.} This can be done by using cold atoms, offering both bosonic and fermionic degrees of freedom with tunable interactions; or by mapping the fermions unitarily and locally to hard-core bosons, which is possible in the presence of a gauge field.
	\item \textbf{In quantum simulators of LGTs  in more than $1+1d$ the gauge field cannot be eliminated, and one must deal with gauge invariance.} This can be done either by imposing gauge invariance on a quantum simulator including both gauge field and matter degrees of freedom (and we have discussed several ways of doing that); by switching, in some cases, to a dual formulation where the redundancy rising from the gauge symmetry is lifted; or, in some cases, by eliminating the matter fields and dealing with a theory including only gauge fields, but with no gauge symmetry. Since the gauge field must be present, one also needs, in some cases, to truncate its local Hilbert spaces, which we have discussed as well.
	\item \textbf{Quantum simulators of LGTs  in more than $1+1d$ have to implement the complicated four-body plaquette interactions.} While such interactions are not "naturally available" in the currently known quantum platforms, we have discussed several ways to obtain them. They can also be avoided when dual formulations are used.
\end{enumerate}

These are special properties of LGTs which cannot be avoided, in general, in more than one space dimension, and have to be taken into consideration when quantum simulators of LGTs in higher dimensions are designed. Since breaking the "bottleneck" of one dimensional simulation is essential from the physical point of view, these questions must be addressed experimentally and revisited theoretically (aiming at further, perhaps more feasible solutions) in the near future.

\section*{Acknowledgement}
This research was supported by the Israel Science Foundation (grant No. 523/20).

\bibliographystyle{ieeetr}
\bibliography{ref}

\end{document}